\documentclass[a4paper,12pt]{article}
\usepackage{latexsym}
\usepackage{graphicx}
\usepackage{bm}
\usepackage{cite}

\def\D{\partial}
\def\EC{\end{center}} 
\def\BC{\begin{center}} 
\def\sfrac#1#2{\hbox{$#1\over#2$}}
\def\BA{\begin{eqnarray}}
\def\EA{\end{eqnarray}}
\def\BAN{\begin{eqnarray*}}
\def\EAN{\end{eqnarray*}}
\def\NN{\nonumber}
\def\BE{\begin{equation}}
\def\EE{\end{equation}}
\def\bnabla{\boldsymbol{\nabla}}
\def\tU{\widetilde U}
\def\tW{\widetilde W}
\def\bU{\overline U}
\def\bV{\overline V}
\def\bW{\overline W}
\def\bZ{\overline Z}
\def\sZ{Z^{\rm s}}
\def\sUz{{U_0^{\rm s}}}
\def\sUu{{U_1^{\rm s}}}
\def\sWz{{W_0^{\rm s}}}
\def\sWu{{W_1^{\rm s}}}
\def\sVu{{V_1^{\rm s}}}

\title{Large scale flow around turbulent spots}
\author{Maher Lagha and Paul Manneville\\
Laboratoire d'Hydrodynamique\\
\'Ecole Polytechnique, F-91128 Palaiseau, France}

\date{Phys. Fluids {\bf19} (2007) 094105.}

\begin{document}

\sloppy

\maketitle

\begin{abstract}
Numerical simulations of a model of plane Couette flow focusing on
its in-plane spatio-temporal properties are used to study the dynamics
of turbulent spots. While the core of a spot is filled with small
scale velocity fluctuations, a large scale flow extending far away and
occupying the full gap between the driving plates is revealed upon
filtering out small scales. It is characterized by streamwise inflow
towards the spot and spanwise outflow from the spot, giving it a
quadrupolar shape. A correction to the base flow is present within the
spot in the form of a spanwise vortex with vorticity opposite in sign to that
of the base flow. The Reynolds stresses are shown to be at the origin
of this recirculation, whereas the quadrupolar shape of the in-plane
flow results from the transport of this recirculation by the base flow
that pumps it towards the spot in the streamwise direction and flushes
it in the spanwise direction to insure mass conservation. These
results shed light on earlier observations in plane Couette flow or
other wall flows experiencing a direct transition to turbulence by
spot nucleation.

\end{abstract}

\maketitle

\section{\label{sec:level1}Introduction}

Being stable against infinitesimal perturbations for all Reynolds
numbers, plane Couette flow (pCf), the shear flow between two parallel
plates moving in opposite directions with velocity $\pm U_{\rm p}$, is
the prototype of flows that require localized finite amplitude
disturbances to be pushed towards a turbulent regime. The transition
is thus characterized by the nucleation and nonlinear growth of
domains of turbulent flow, separated from laminar flow by sharp
fronts and called {\it turbulent spots\/} 
(e.g., \cite{TA92,DHB92,DD94,DD95,Ti95}).
This kind of transition is not restricted to pCf but is also present
in plane Blasius (boundary layer) flow~\cite{Em51,GBR81} or plane
Poiseuille flow \cite{CWP82}. A review of some relevant laboratory
experiments is given by Henningson {\it et al.} \cite{Hetal94} and of
their numerical counterpart given by Mathew \& Das \cite{MD00}.
In practice, direct transition to turbulence {\it via\/} spots can be
expected whenever no low-Reynolds number instability of inertial
origin exists, whereas turbulent 
solutions to the Navier--Stokes equations may exist and compete with
the laminar base flow at moderate Reynolds number \cite[Chap.6, \S6.3]{Ma04a}.

Growing turbulent spots in pCf have been studied both
experimentally \cite{TA92,DHB92,DD94,DD95,Ti95} and numerically
\cite{LJ91,DM01,SE01}.
In their pioneering direct simulations of Navier--Stokes equations with
realistic no-slip boundary conditions, Lundbladh \& Johansson
\cite{LJ91} pointed out that (i) the wall-normal velocity component
---typical of internal irregular small scale structures--- faded away
outside the spot but (ii) slowly varying in-plane velocity  
components extended far outside with an inwards streamwise motion 
towards the spot at the streamwise edges and an outward spanwise
motion at its spanwise edges. These observations were made
by low-pass Gaussian filtering the small scales of the velocity
field at mid-gap. Tillmark~\cite{Ti95} confirmed them
experimentally by detecting the outwards spanwise component that
developed over the full gap between the plates.

More recently, Schumacher \& Eckhardt \cite{SE01} re-investigated the
growth of turbulent spots by means of direct numerical simulations but
using unrealistic free-slip boundary conditions at the plates. By
averaging the flow field between the two plates, they also observed that
the turbulent spot was accompanied by an overall spanwise outflow and
streamwise inflow, which they termed {\it quadrupolar\/}.

Spots seem to behave as obstacles in the base flow \cite{GBR81,LW89,DD94}.
Accordingly, they introduce additional pressure fields induced
by the distribution of Reynolds stresses associated with the small
scale fluctuations inside the spot and generating the large
scale flows. A similar interpretation was put forward by
Hayot \& Pomeau \cite{HP94} who introduced a {\it back-flow\/} to
explain the organization of spiral turbulence in cylindrical Couette
flow \cite{Co65}, with possible application to the banded turbulent 
regime discovered more recently in pCf \cite{Petal02}
and numerically studied by Barkley \& Tuckerman \cite{BT07}.

Previous experimental studies by Bottin {\it et al.}  \cite{Betal98}
have shown that, in the lowest part of the 
transitional Reynolds number range,  flow patterns of interest
extend over the full gap. We take advantage of this
observation to study the dynamics of spots using numerical simulations
of a previously derived model of pCf shown to display sufficiently
good properties for this purpose \cite{La06}. The model is sketched in
\S\ref{SII} and completed in the Appendix. Typical
results of 
simulations are presented in \S\ref{SIII} emphasizing the output of
the filtering 
procedure: (i) the in-plane quadrupolar flow outside the spot and (ii)
a spanwise recirculation cell inside. These observations are then
interpreted in \S\ref{SIV} where the generation of these two large
scale flow components is explained in terms of Reynolds stresses
averaged over the surface of the spot. In the concluding section, we
summarize our results and point to their relevance to the
interpretation of previous observations in other wall flows of less
academical interest, such as plane Poiseuille \cite{HK91} or Blasius
flows \cite{SK04}.

\section{The model}
\label{SII}

The model used here is an extension to realistic {\it no-slip\/} boundary
conditions of an earlier model proposed by one of us \cite{ML00} for
unrealistic {\it free-slip\/} boundary conditions.
It is derived in \cite{La06} from the Navier--Stokes
equations through a systematic Galerkin method involving expansions
in terms of polynomials, functions of the cross-stream coordinate $y$
multiplied by amplitudes describing the in-plane ($x,z$) space
dependence of the full velocity field. 
The equations are written for the perturbation $(u',v',w',p')$ to the
laminar basic flow $U_{\rm b}{\bf \hat x}$, where $\bf\hat x$ denotes the
streamwise direction, i.e. $u=U_{\rm b}(y)+u'$; $v'$ and $w'$ denote the
perturbations in the cross-stream and spanwise directions,
respectively, $p'$ being the pressure perturbation.
Lengths are scaled by the half-gap between the plates
$h$, and velocities by $U_{\rm p}$ so that the time scale is
$h/U_{\rm p}$. The control parameter is the Reynolds number defined 
as $R=U_{\rm p} h/\nu$, where $\nu$ is the fluid's kinematic viscosity,
and the dimensionless base flow
profile reads $U_{\rm b}(y)=y$ for $y\in[-1,1]$.

In accordance with experimental observations \cite{Betal98}, 
truncation of the Galerkin expansion at lowest consistent order
is performed, reducing the set of basis functions to:
\BA
\label{pertur}
u' (x,z,t,y)&=&U_0(x,z,t)B(1-y^2)+U_1(x,z,t)Cy(1-y^2),\\
\label{pertur2}
v'(x,z,t,y)&=&V_1(x,z,t) A(1-y^2)^2,\\
\label{pertur3}
w'(x,z,t,y)&=&W_0(x,z,t)B(1-y^2) +W_1(x,z,t)Cy(1-y^2),
\EA
where $A$, $B$, and $C$ are normalisation constants.
These expressions are inserted in the continuity and Navier--Stokes
equations, and projections of the results on the same basis functions
using the canonical scalar product
$\langle f,g\rangle=\int_{-1}^{+1} f(y)g(y)\,{\rm d}y$, are performed,
which yields a set of coupled partial differential equations. For
example, the projection of the continuity equation reads:
\BE
\label{E-cont}
\D_x U_0+\D_z W_0=0\,, \quad\quad \D_x U_1+\D_z W_1=\beta \, V_1\,,
\EE
with $\beta=\sqrt{3}$. The complete model is given in the Appendix.
Here we only display the equation for the amplitude $U_0$
of the streamwise velocity component which is even in $y$:
\BE
\D_t U_0 +N_{U_0}=-\D_x P_0 -a_1\D_x U_1 -a_2 V_1
+R^{-1} \left(\Delta -\gamma_0\right)U_0\,,\label{E-U0ns}
\EE
where $\Delta=\D_{xx}+\D_{zz}$ and with:
\BE
N_{U_0}=\alpha_1(U_0\D_xU_0+W_0\D_zU_0)+
\sfrac12\alpha_2(U_1\D_xU_1+V_1\beta'U_1+W_1\D_zW_1)\,,
\label{E-NLU0}
\EE
just to show that each equation has the form expected for a
hydrodynamic problem. In particular, nonlinearities have the same structure  
as the classical advection term $\mathbf v\cdot\bnabla\mathbf v$. In
the same way, the last term in (\ref{E-U0ns}), with the factor
$R^{-1}$, accounts for the viscous dissipation associated with the
cross-stream parabolic ($\gamma_0$) and in-plane dependencies of $U_0$. 
This flow component can straightforwardly be identified as the
{\it streamwise\/} streak amplitude, so that the source term $-a_2V_1$ on 
the r.h.s. of (\ref{E-U0ns}) accounts for the {\it lift-up\/}
mechanism since $V_1$ is the cross-stream velocity fluctuation. 
The physical role of the linear term $-a_1\D_x U_1$
will be considered later.

On general grounds, the Reynolds--Orr equation governs the perturbation
energy $E(t)=\frac{1}{2}\int_{\mathcal V} (u'^2+v'^2+w'^2)\, d\mathcal
V$, where $\mathcal V$ is the volume of the domain. It can be
symbolically written as $\frac{\rm d}{{\rm d} t} E=P-D$, where $P$ is the
energy production issued from the interaction of the perturbation with
the base flow $U_{\rm b}(y)\equiv y$, $P=-\int_\mathcal V u'v' \frac{\rm
  d}{{\rm d}y}U_{\rm b}\, d\mathcal V$, and $D$ is the dissipation due
to viscous effects. In our model, one readily gets
$P=-\int_{\mathcal S} \chi U_0 V_1\,d\mathcal S$, where $\mathcal S$
is the surface of the domain and $\chi$ is a positive
constant. Since $V_1$ generates $U_0$ through the lift-up mechanism,
regions where the Reynolds stress $-U_0 V_1$ is positive, thus
destabilizing the base flow and contributing to the turbulence production,
are those with $U_0>0$ and $V_1<0$ or the reverse, which
obviously correspond to $Q_2$ and $Q_4$ events identified in the
literature, see for example \cite{Pa01}.

The main limitation of the model comes from its low order
truncation. In fact, expressions (\ref{pertur}--\ref{pertur3}) are
only the first terms of series expansions and the derivation of
models truncated at higher and higher orders remain possible. Up to
now, this has not been done for several converging reasons, the main
ones being that ({\it i\/}) the Reynolds number range we are interested in
corresponds to the lower part of the pCf's transitional regime where
departures from laminar flow are known to occupy the full gap \cite{Betal98},
({\it ii\/}) $U_1$ already contains the lowest order non-trivial
correction to the base flow thought to be important in the discussion
of the laminar--turbulent coexistence \cite{HP94}. Accordingly we
believe that the lowest order model is 
sufficient to account for the large scale features present in the
experiment at least at a qualitative level, the alternative being to
turn to direct numerical simulations and not to consider a much more
cumbersome higher order model. The discussion in \S\ref{SIV} supports
the validity our approach {\it a posteriori\/}.

\section{Numerical simulations of turbulent spots}
\label{SIII} 

Our model was integrated on a rectangular $(x,z)$ domain with
periodic boundary conditions, while being written for stream-functions
$\Psi_0$, $\Psi_1$ and velocity potential $\Phi_1$ related to the
velocity amplitudes through:
\BA
\label{UW0}
&&\hspace*{-1em}U_0=\tU_0- \D_z \Psi_0\,, \quad W_0=\tW_0+\D_x
\Psi_0\,,\\
\label{UVW1}
&&\hspace*{-1em}U_1=\tU_1+ \D_x \Phi_1-\D_z \Psi_1\,,
\  W_1=\tW_1+\D_z \Phi_1+\D_x \Psi_1\,,\ 
V_1=\Delta \Phi_1/\beta\,.
\EA
A standard, Fourier based, pseudo-spectral
code was implemented with nonlinear terms and linear non-diagonal
terms (e.g. $-a_1\D_x U_1 -a_2 V_1$ in (\ref{E-U0ns})) evaluated in
physical space and integrated in time using a second order
Adams--Bashforth scheme. The
necessary introduction of $\tU_0$,\dots\ is commented upon in the
Appendix. Simulations were performed in a  domain of
size $(L_x\times L_z)=(128\times128)$ with effective space steps 
$\delta x=\delta z=0.25$ and $\delta t=0.01$. These values were
retained as a good compromise between accuracy and the possibility to
let sufficiently wide systems evolve over sufficiently large periods
of time \cite{La06}. Concerning the accuracy problem, it should be
noted that small-scale in-plane structures are pieces of streaks and
streamwise vortices with typical size larger than 3, which 
makes more than 10 collocation points {\it per\/} structure. Smaller
time steps did not produce results different from those shown here
during comparable time lengths.

As an initial condition, we took localized expressions for $\Psi_0$,
$\Psi_1$, and $\Phi_1$:
$$
\Psi_0(x,z,t=0)=\Psi_1(x,z,t=0)=\Phi_1(x,z,t=0)=A\exp^{-(x^2+z^2)/S}\,
$$
where $A$ is an amplitude and $S$ is the size of the
germ. Parameters $A=5$ and $S=2$ were found efficient in
generating turbulent spots for $R=250$, well beyond $R_{\rm g}\sim
173$, above which sustained turbulence is expected in our model
\cite{La06}. In practice,  due to the highly unstable
characteristics of the flow at such values of $R$, the apparent
simplicity of the initial condition played no role after a few time units.

\begin{figure}
\begin{center}
\includegraphics[width=0.5\textwidth,clip]{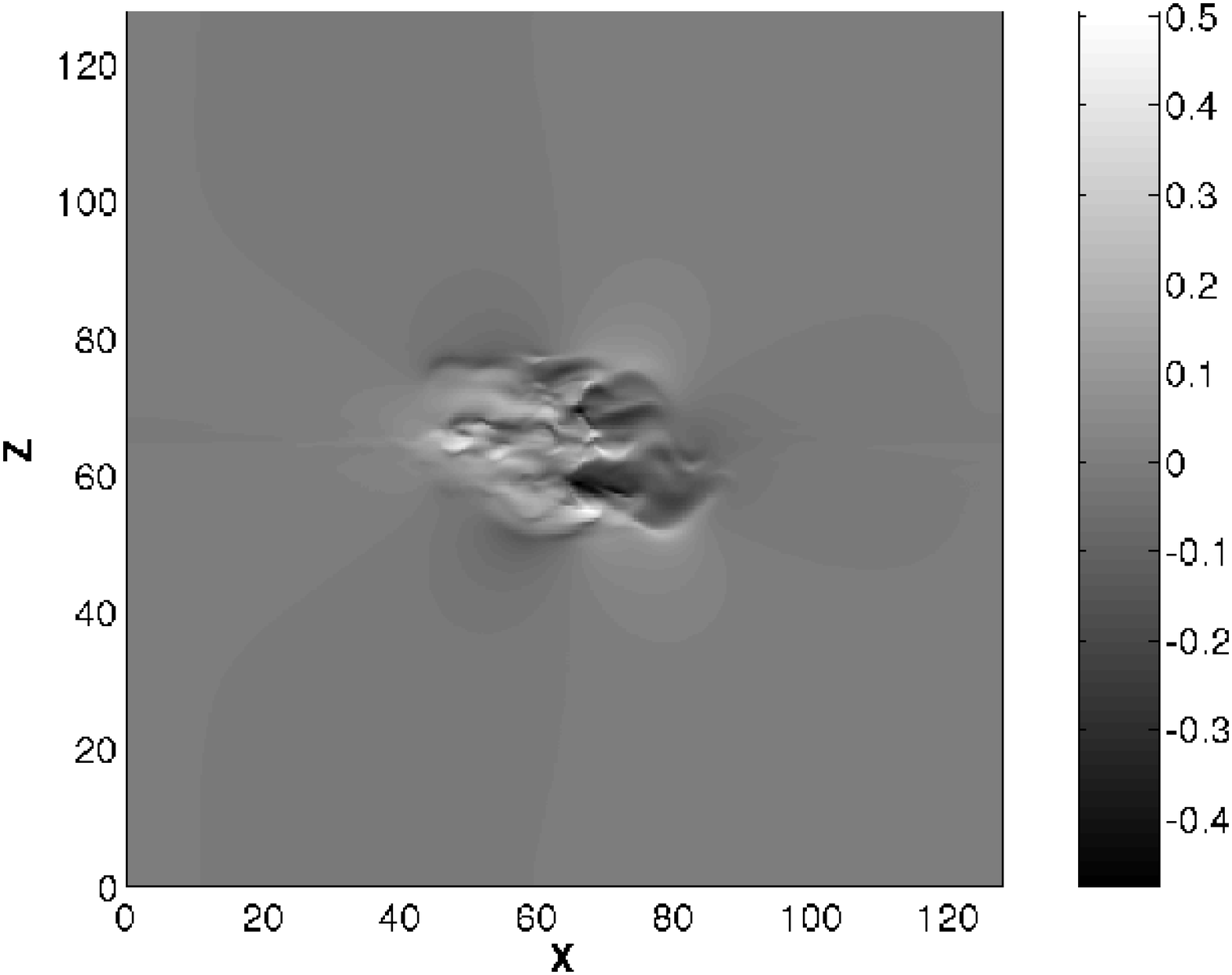}\hfill
\includegraphics[width=0.5\textwidth,clip]{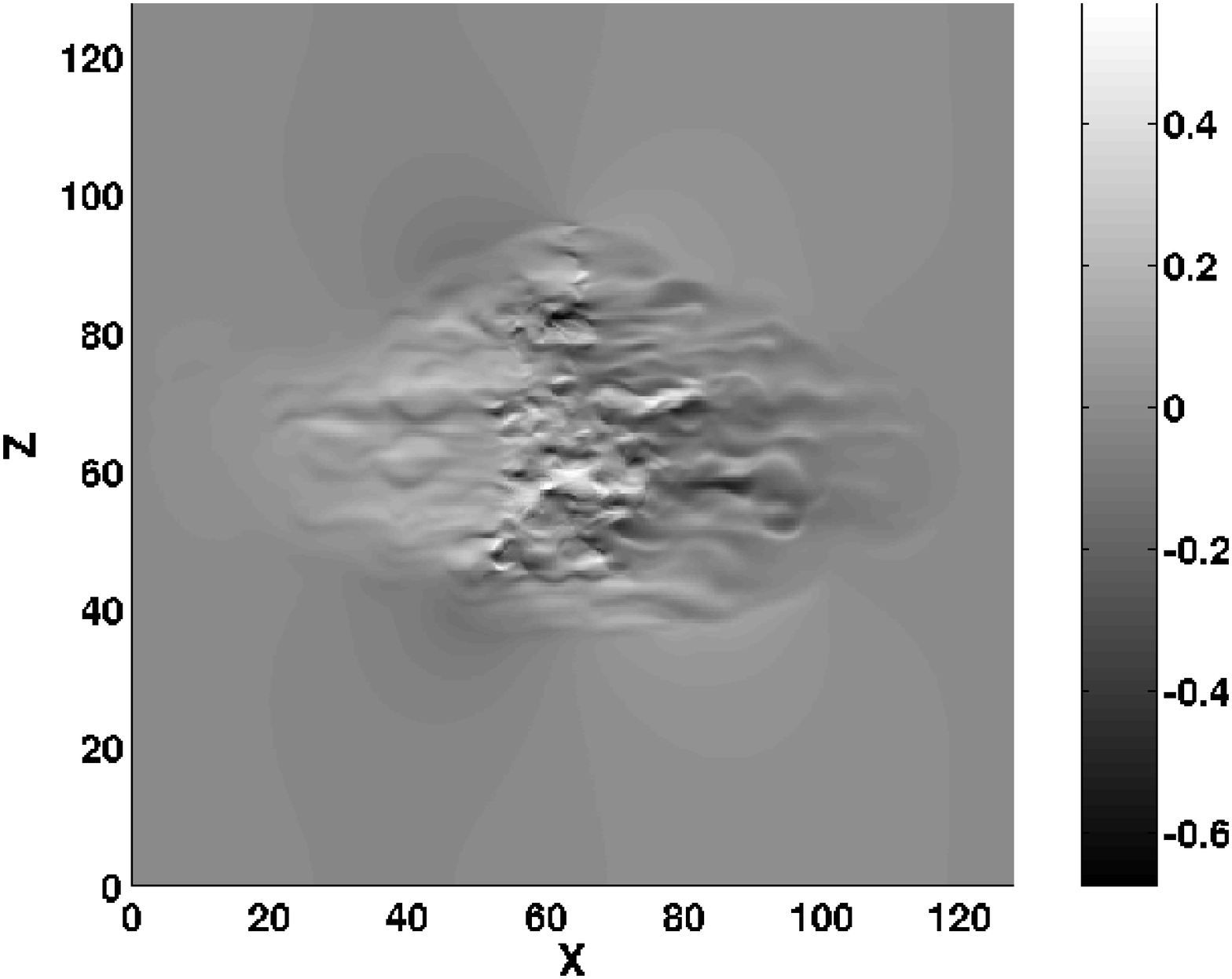}\hfill
\includegraphics[width=0.5\textwidth,clip]{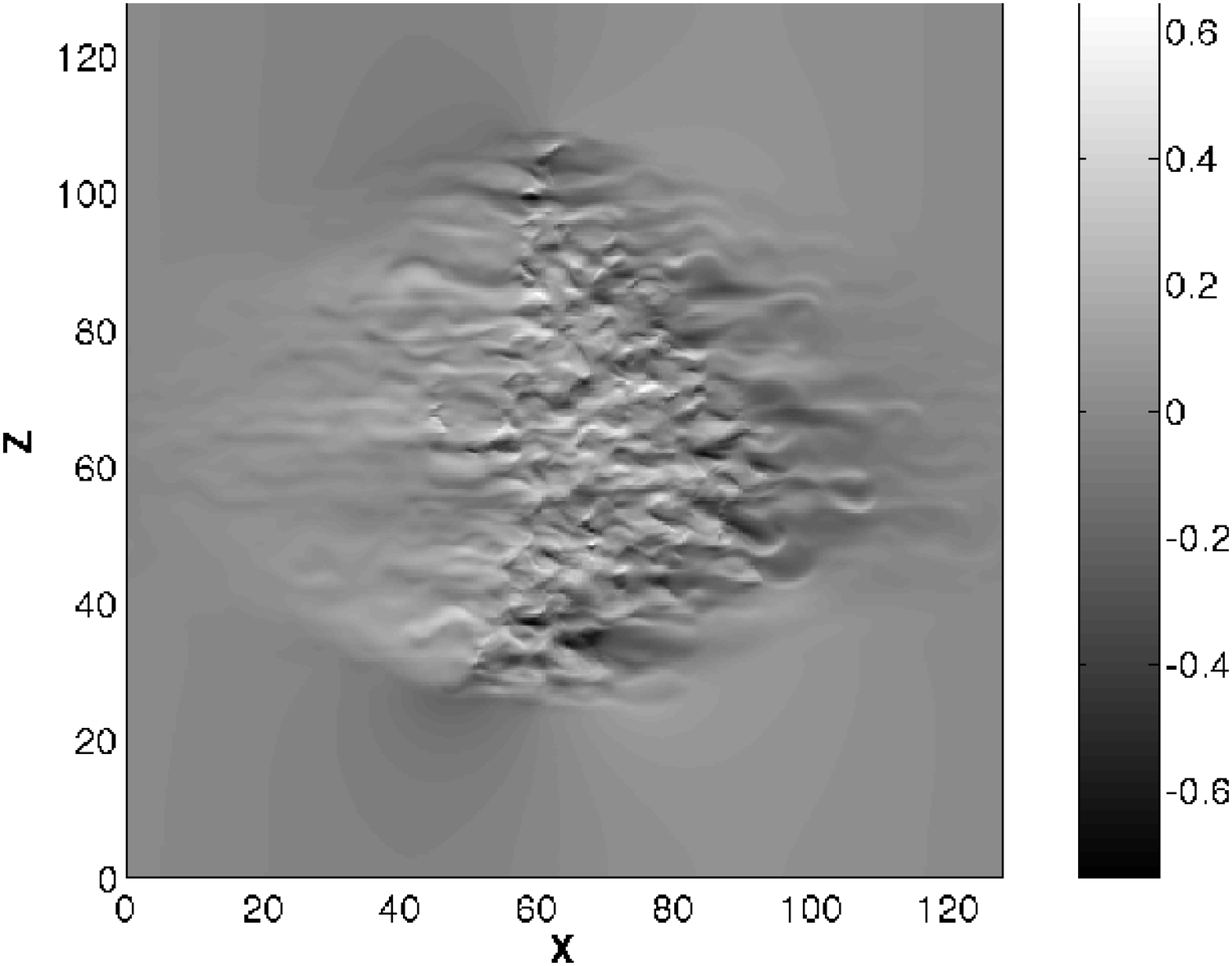}\hfill
\includegraphics[width=0.5\textwidth,clip]{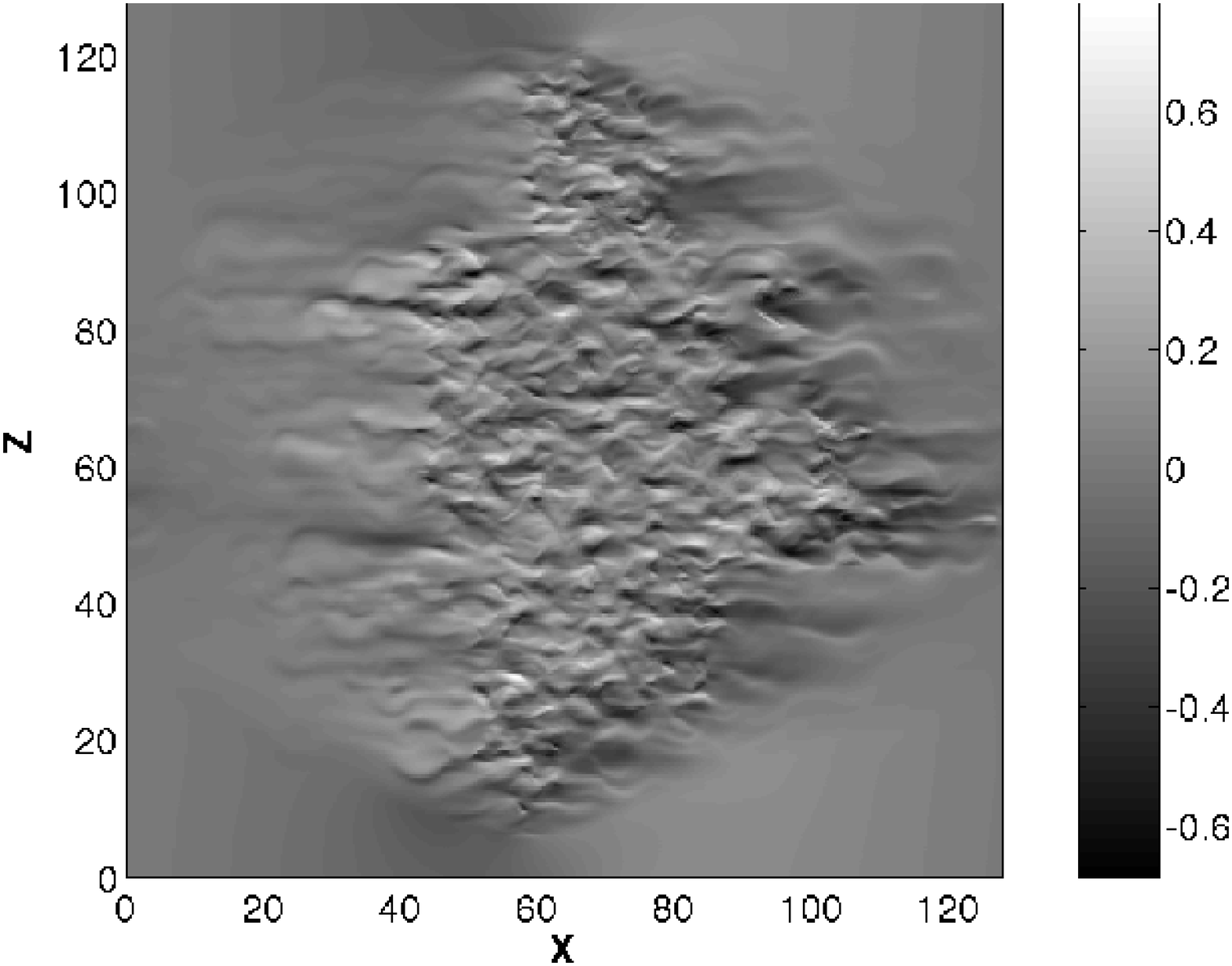}
\caption{Growth of a turbulent spot at $R=250$ in a wide domain ($L_x\times
  L_z=128\times 128$). Field of amplitude $U_0(x,z,t)$ in gray levels at
  $t=50$, $150$, $250$ and $350$ (from left to right and top to
  bottom). The whole domain becomes uniformly turbulent
  at $t\approx 700$.}
\label{spot_u0}
\end{center}
\end{figure}

Spots are best illustrated by their most spectacular feature, namely
their streamwise streaky structure \cite{TA92,DHB92,GBR81,CWP82}. In
turn, the latter is best visualized from the amplitude $U_0$ since
streamwise streaks are easily identified as regions where
$|W_0|\ll|U_0|$ alternating in the spanwise direction, and since
$U_0$ is associated to velocity perturbations that are maximum in the
mid-gap plane $y=0$. Figure~\ref{spot_u0} displays gray-level
snapshots of $U_0$ at different times after launching.
Denoting by  $(x_{\rm C},z_{\rm C})$  the in-plane coordinates of the
center of the spot we see that, contrasting with the cases of plane
Poiseuille or boundary layer flows, the spot does not drift due to the
absence of mean advection. One can also notice 
its overall ovoid shape with dominant negative values (dark gray) for 
$x>x_{\rm C}$ and positive values (light gray) for $x<x_{\rm C}$. 
Regions where $U_0$ is positive correspond to high and low speed streaks
for $y>0$ and $y<0$, respectively, which compares
well with the experimental observations in \cite{Betal98}.

\begin{figure}
\begin{center}
\includegraphics[width=0.5\textwidth,clip]{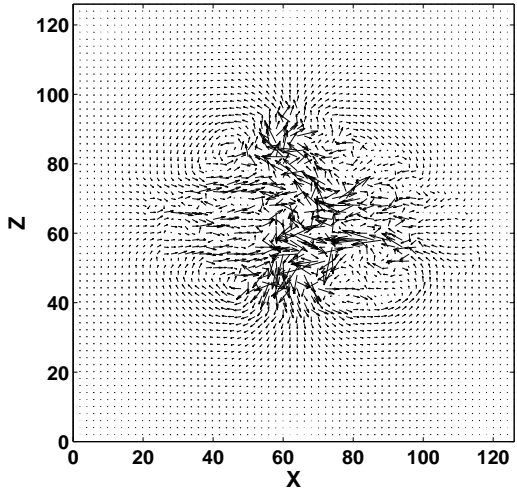}
\caption{Streak flow field $(U_0,W_0)$ at $t=150$.}
\label{spot_u0b}
\end{center}
\end{figure}

In the sequel, we study the state at $t=150$ but results and
conclusions are identical at different times. The complete field
$(U_0,W_0)$ corresponding to this reference state is displayed in
Figure~\ref{spot_u0b}. Except in the very center of the spot that
looks rather messy, streamwise structures are easily recognized but
the trace of the large scale quadrupolar flow, of main concern in the
present paper, is already visible.

\begin{figure}
\begin{center}
\includegraphics[width=0.5\textwidth,clip]{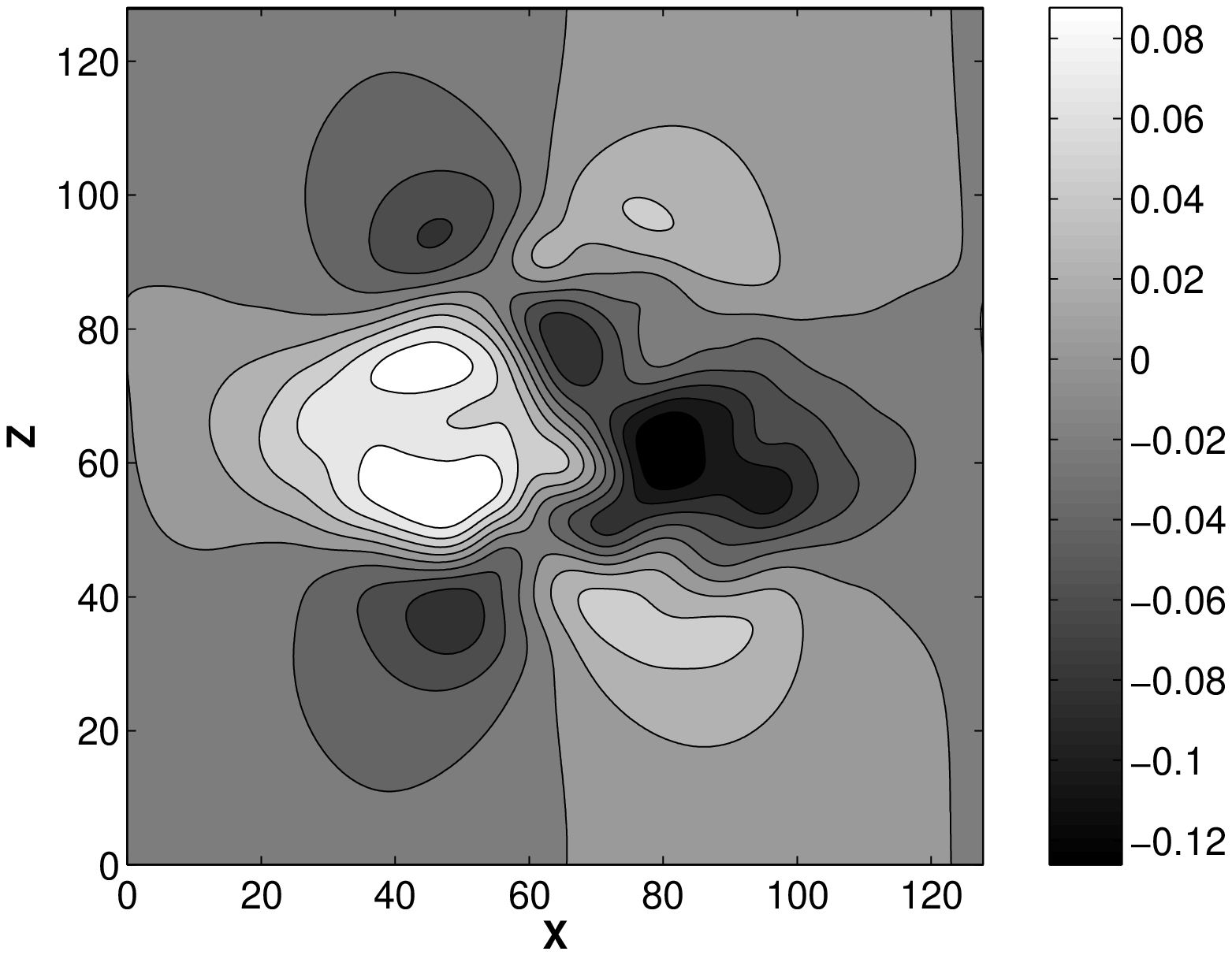}\hfill
\includegraphics[width=0.5\textwidth,clip]{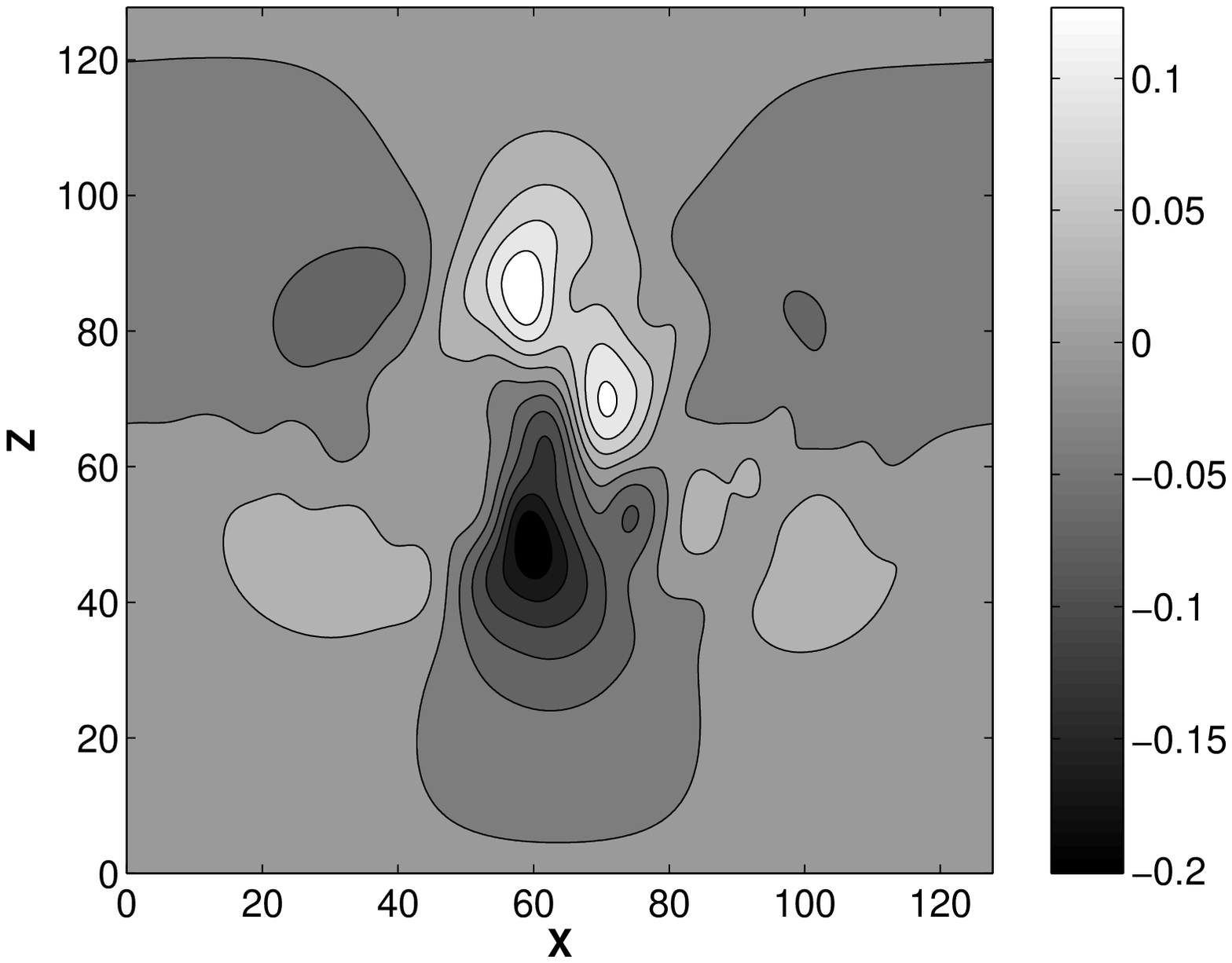}

\includegraphics[width=0.5\textwidth,clip]{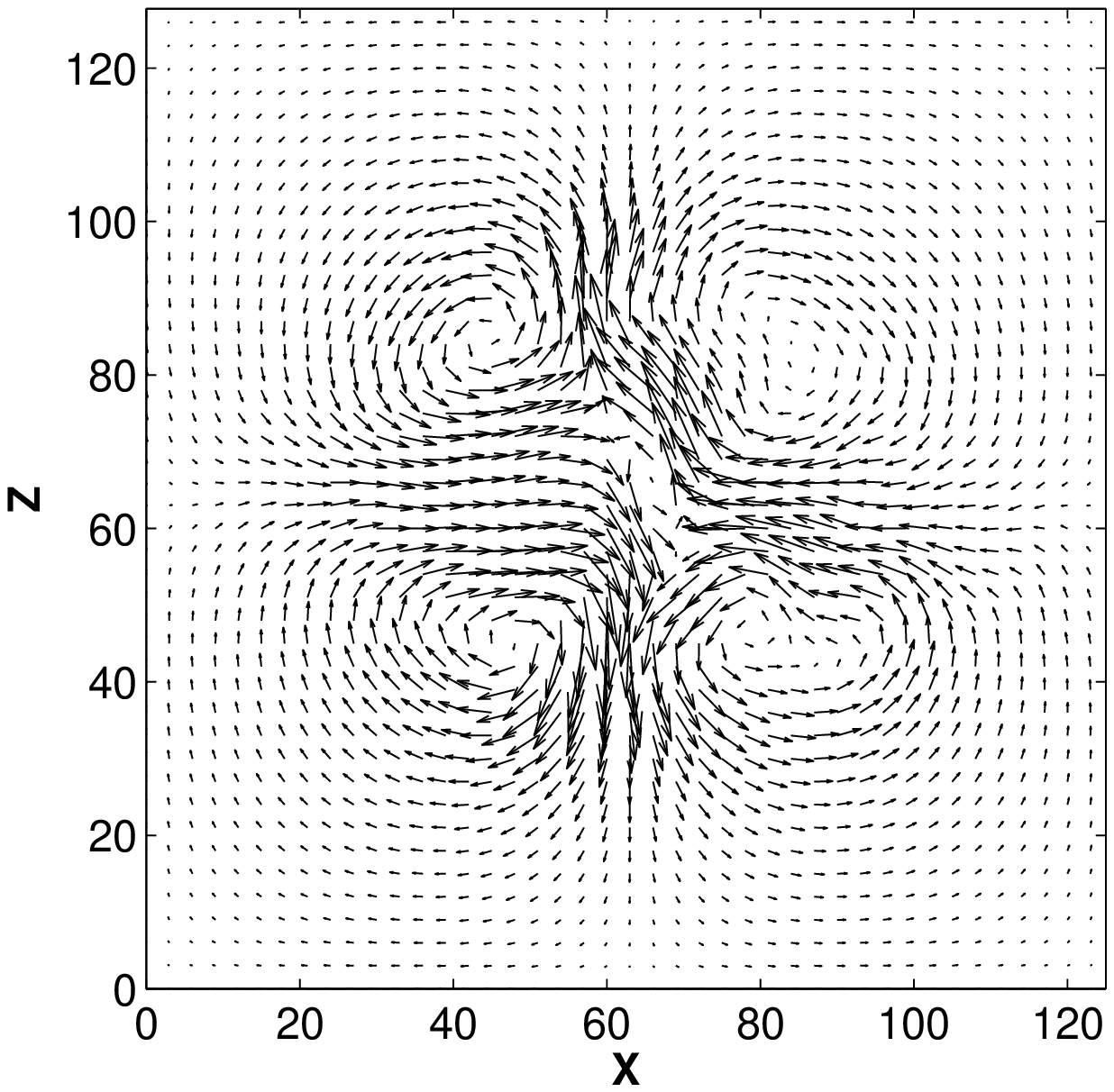}\hfill
\includegraphics[width=0.5\textwidth,clip]{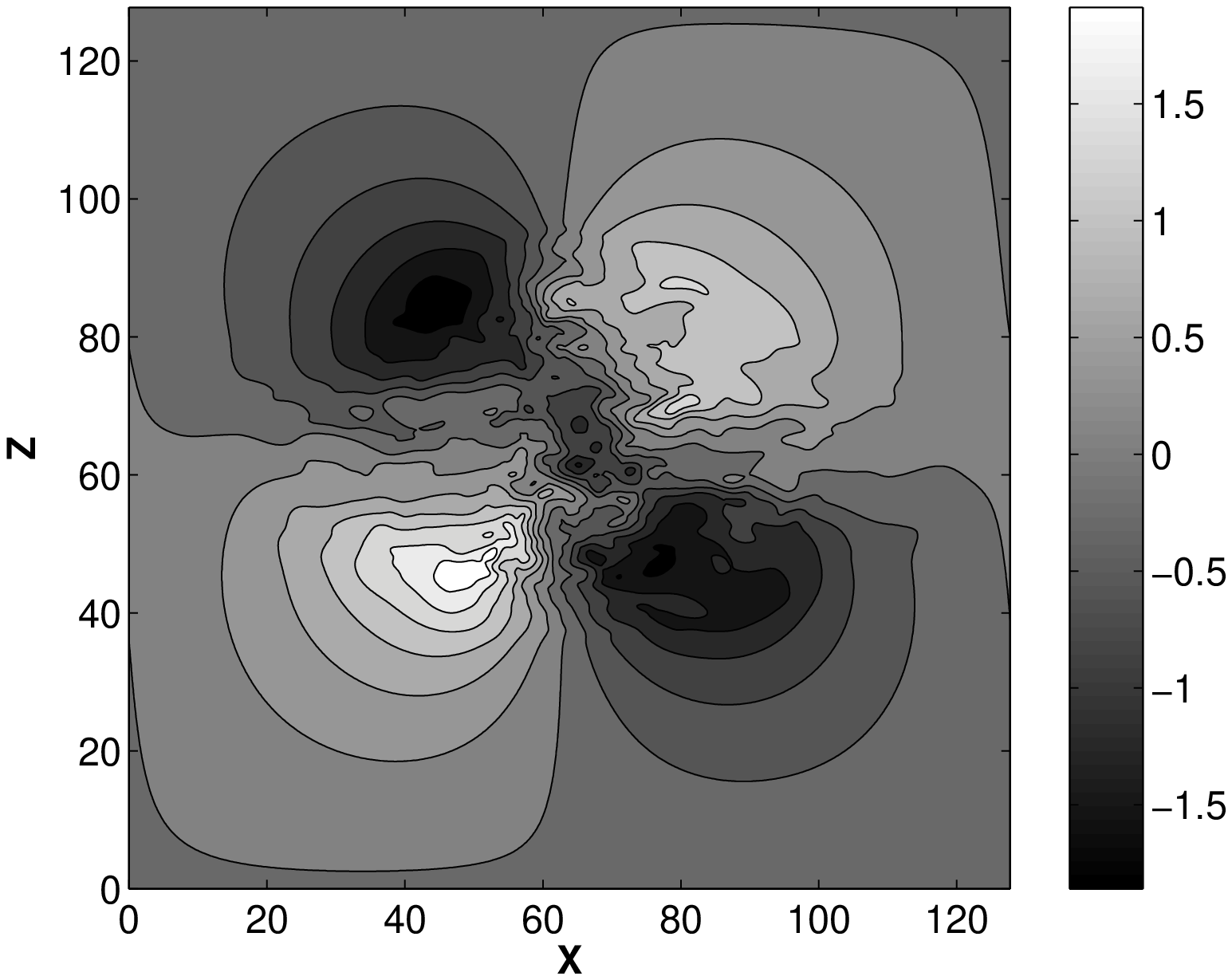}

\caption{Top: level lines of averaged velocity components $\bU_0$ (left) and
  $\bW_0$ (right), illustrating large scale streamwise
  inflow and spanwise outflow around the spot. Bottom, left:
  representation of this flow as vectors. Bottom, right: level lines
  of the unfiltered stream function $\Psi_0$.}
\label{ob2}
\end{center}
\end{figure}

As done by Lundbladh \& Johansson \cite{LJ91}, we now proceed to
the elimination of small scales using a Gaussian filter in spectral space:
\BE
\label{E-Gauss}
\widehat{\bZ}(k_x,k_z)=\widehat Z(k_x,k_z)
\exp[-(k_x^2+k_z^2)/(2\sigma)^2]\,,
\EE
where the hat denotes the Fourier transform of any quantity $Z=U_0$,\dots.
In physical space, this corresponds to a convolution with a kernel
$\propto \exp\left(-\sigma\sqrt{\zeta_x^2+\zeta_z^2}\right)$ where $\sigma$ is the
parameter controlling the width of the domain over which the small
scales are smoothed out by the operation. Small scales, indicated by
superscript `s', are recovered afterwards from the relation $\sZ=Z-\bZ$.

The diameter of the Gaussian averaging window has to be chosen in
accordance with the size of the modulations to be eliminated, here the
small scale streaks with spanwise wavelengths of the order $3$--$6$ as
can be guessed from Figure~\ref{spot_u0b}. We used
$\sigma=\pi/11$, but the results were found to be rather insensitive to
this choice provided that $\sigma$ is sufficiently small.

As seen in Figure~\ref{ob2}, this filtering procedure yields a clear
picture of the flow outside the spot: the overall pattern formed by the
in-plane components $\bU_0$  and $\bW_0$ has a quadrupolar aspect that
could already be guessed from the consideration of the unfiltered
stream-function $\Psi_0$ whose Laplacian is related to its vortical
contents. In what follows, we term {\it drift flow\/} the large-scale
velocity field $(\bU_0,\bW_0)$ with Poiseuille-like cross-stream
profile by analogy with the case of Rayleigh--B\'enard convection
where a flow with the same global features was introduced by Siggia \&
Zippelius \cite{SZ81}.

\begin{figure}
\begin{center}
\includegraphics[width=0.5\textwidth,clip]{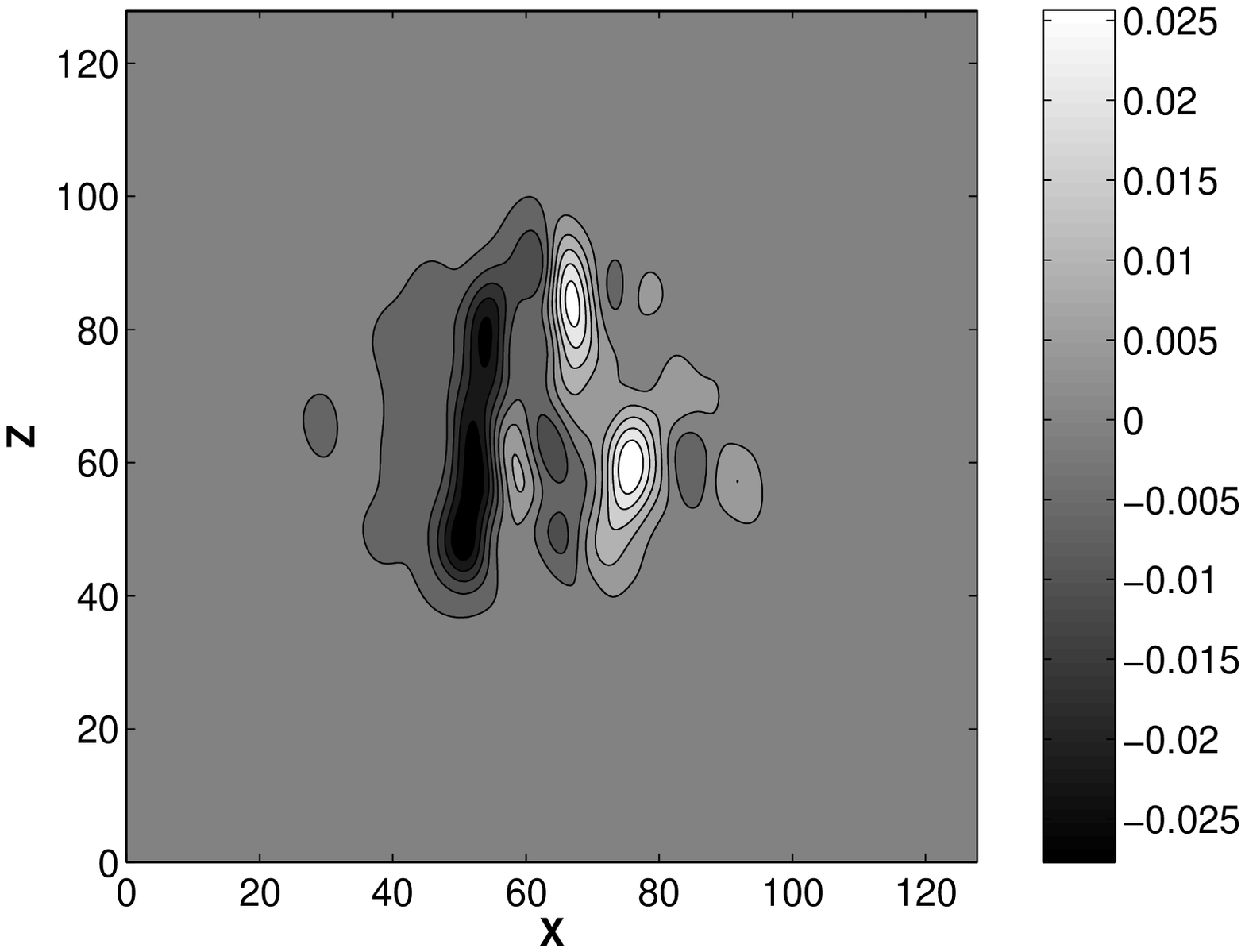}\hfill
\includegraphics[width=0.4\textwidth,clip]{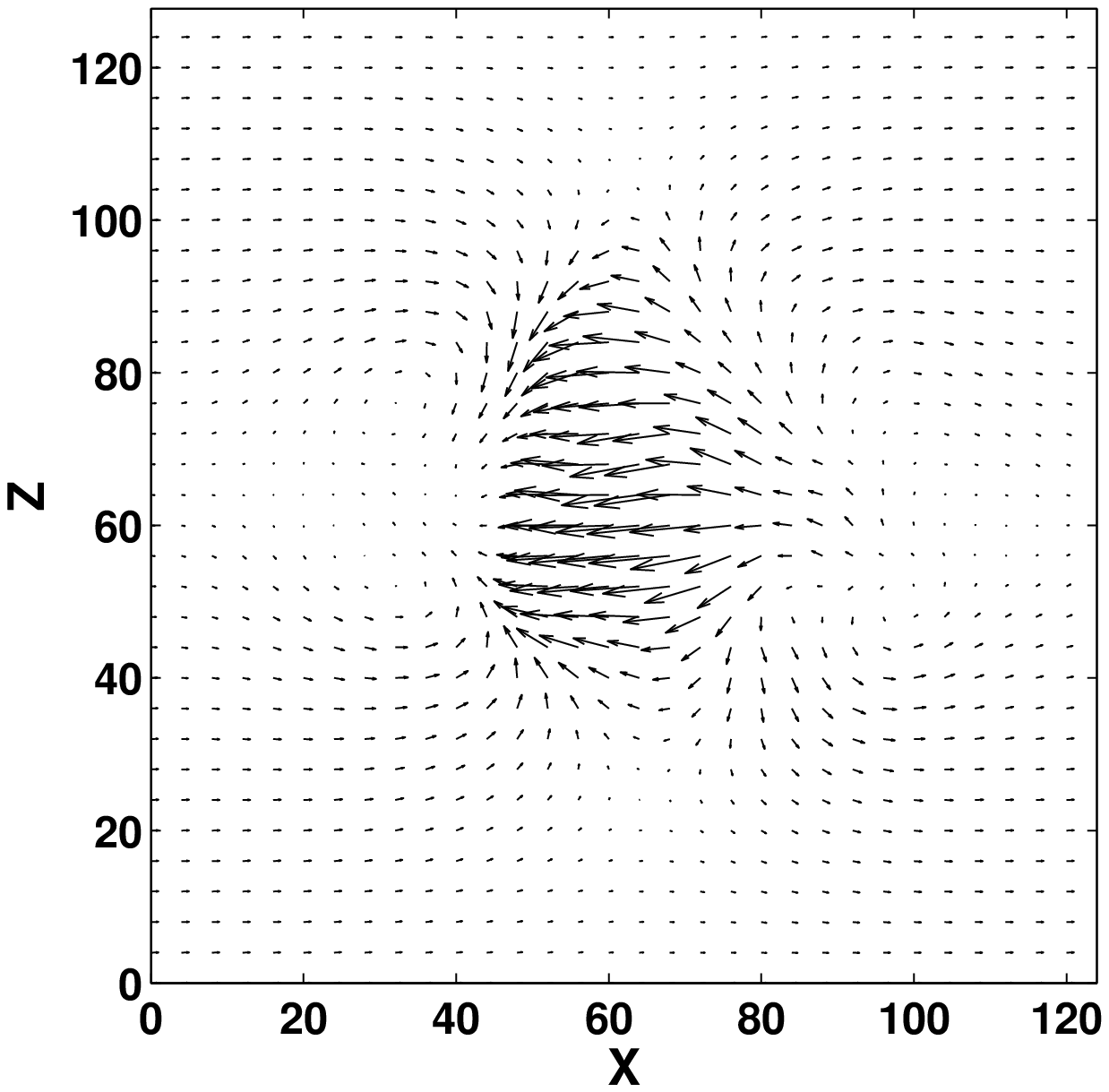}
\caption{Velocity amplitudes $\bV_1$ (left) and $(\bU_1,\bW_1)$ (right).}
\label{ob4}
\end{center}
\end{figure}

Figure~\ref{ob4} displays the velocity components associated to the
fields $\Psi_1,\Phi_1$. The distribution of the amplitude of
$\bV_1$, displayed in the left panel, represents an average wall-normal
motion which is maximum in the mid-plane $y=0$, positive on the right
of the spot's center $x>x_{\rm C}$ and negative on its left. In turn,
the flow $(\bU_1,\bW_1)$ shown in the right panel consists in a region
centered around the spot where $|\bU_1|\gg |\bW_1|$ and
$\bU_1<0$. This structure is easily interpreted as a wide spanwise
recirculation cell with vorticity opposite in sign to that of the base
flow. It is further reminiscent of what can be deduced from  DNS
simulations of Lundbladh and Johansson \cite{LJ91}, as displayed in
seen their Fig.~9.

\begin{figure}
\begin{center}
\includegraphics[width=0.4\textwidth,clip]{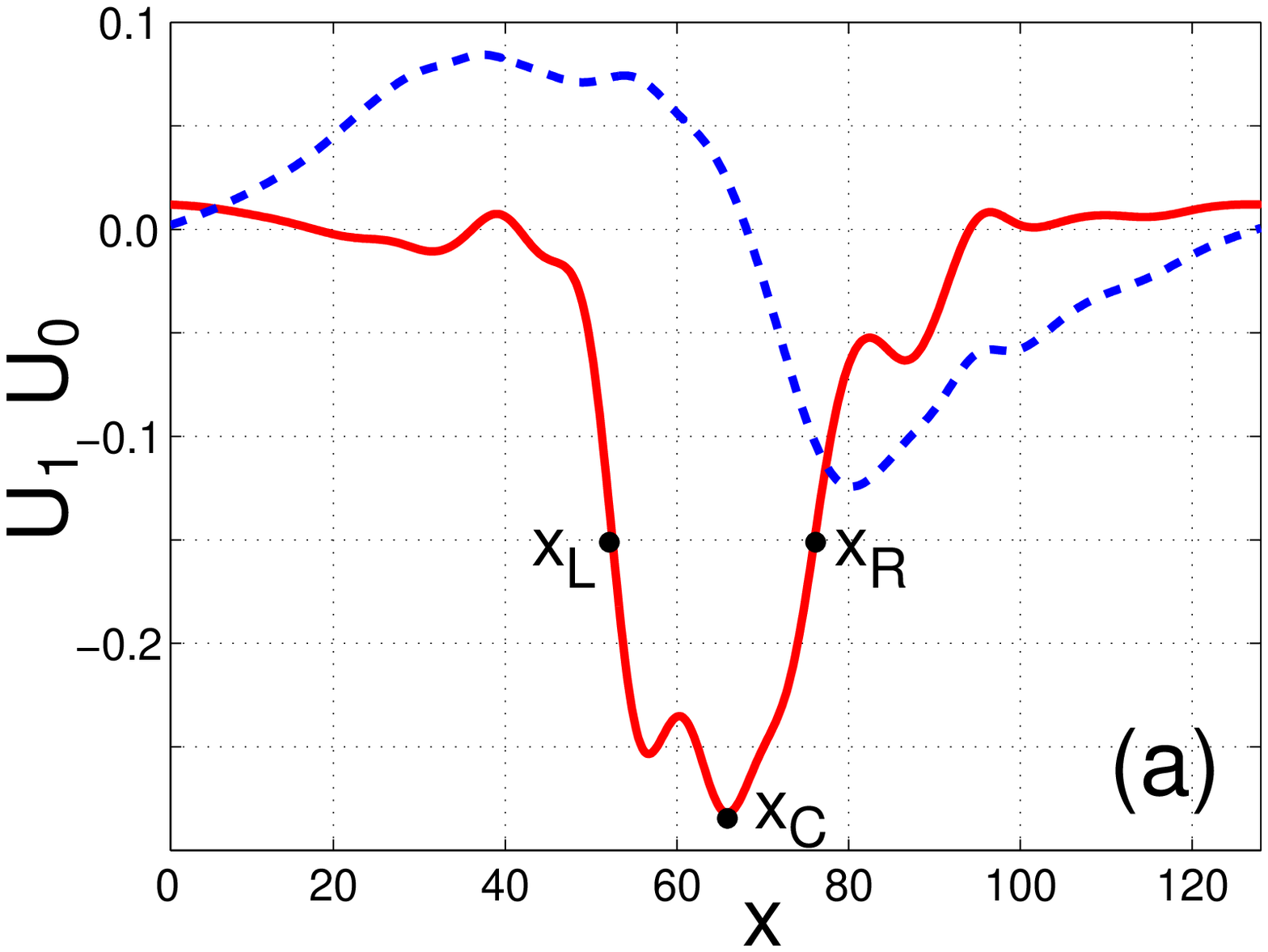}
\hskip0.05\textwidth
\includegraphics[width=0.4\textwidth,clip]{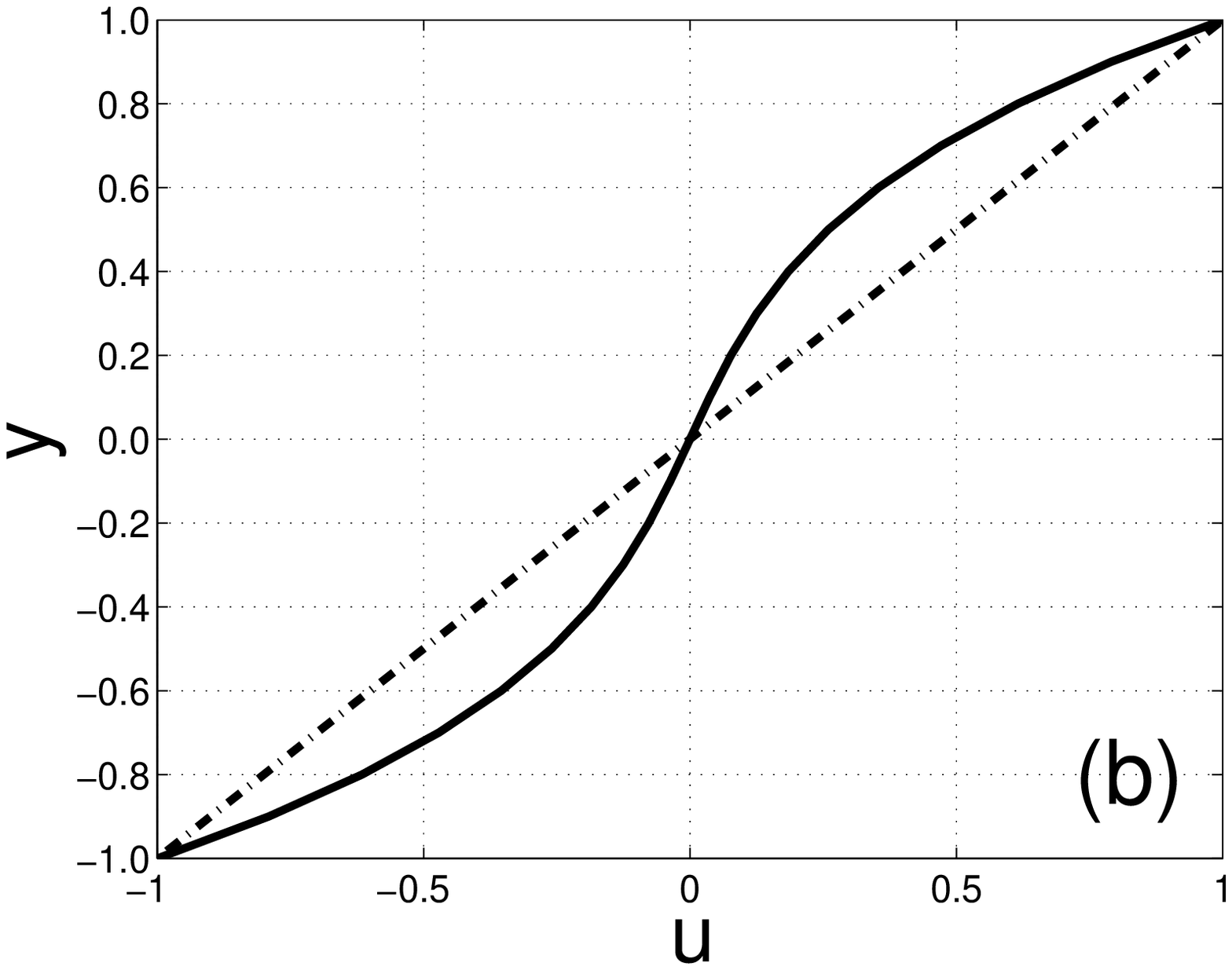}

\includegraphics[width=0.4\textwidth,clip]{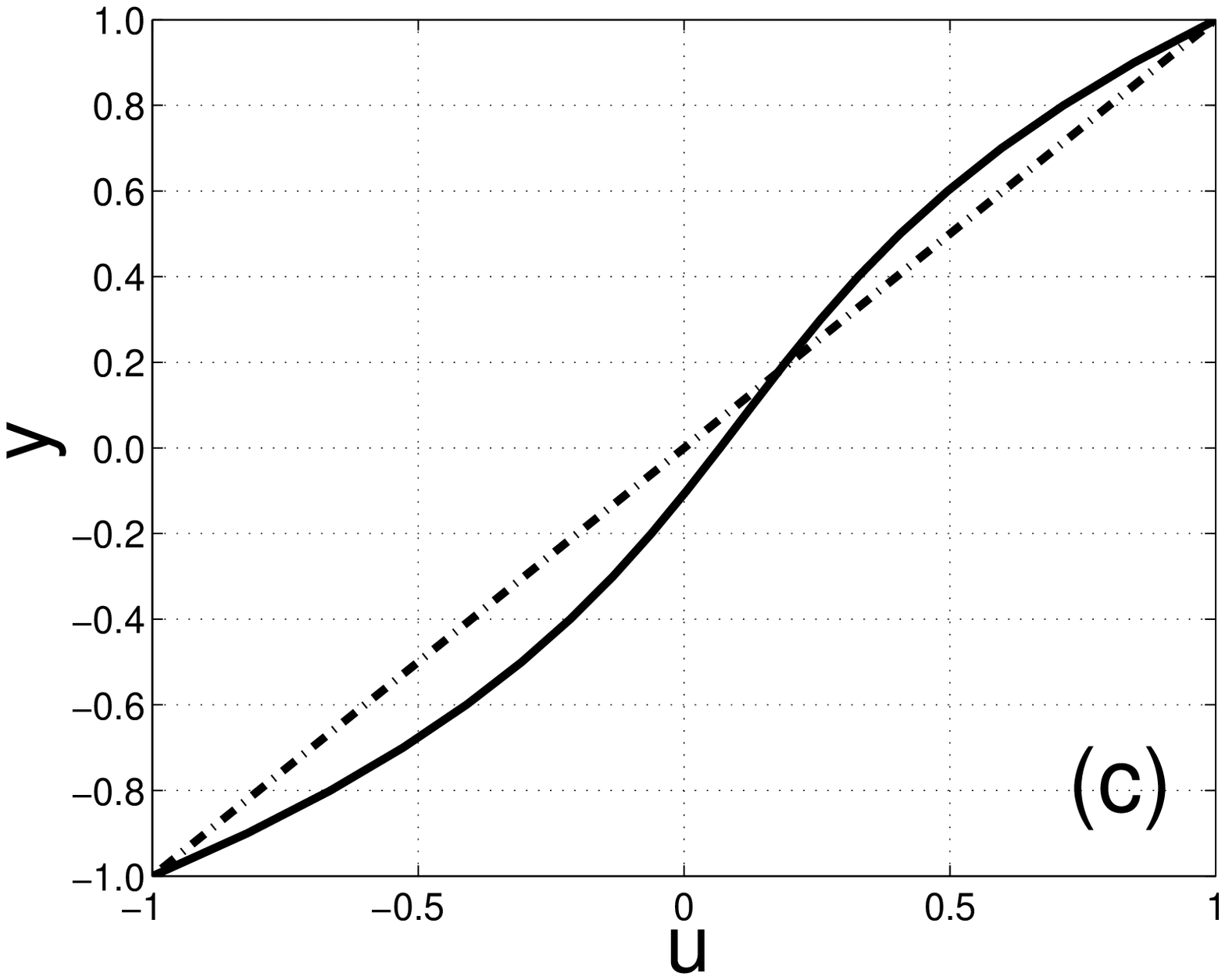}
\hskip0.05\textwidth
\includegraphics[width=0.4\textwidth,clip]{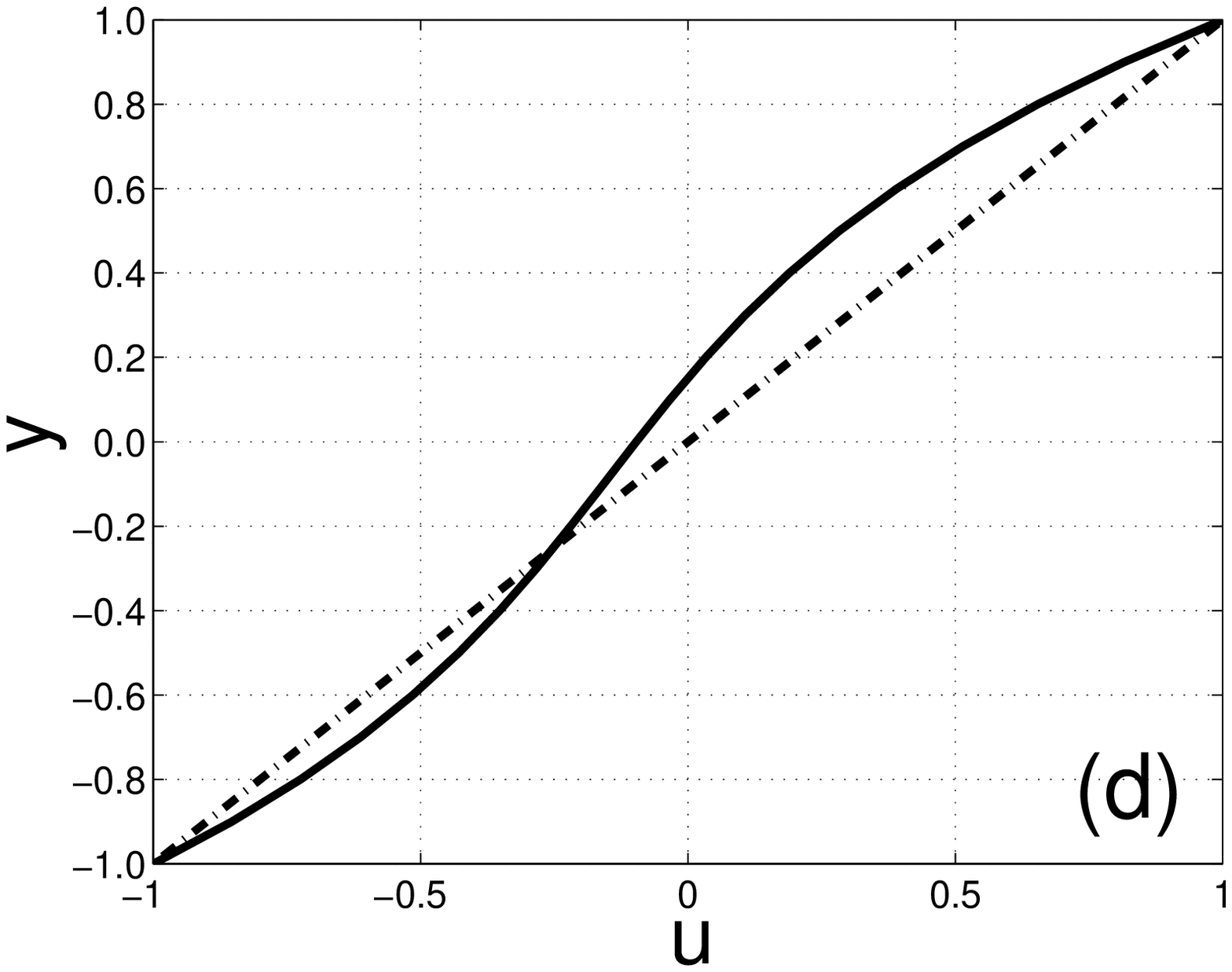} 
\caption{(a) $\bU_1$ (red-solid) and $\bU_0$ (blue-dashed) as
  functions of coordinate $x$ along the streamwise center-line. (b--d)
  Full average streamwise velocity profiles $\overline U(y)$ at
  $x=x_{\rm C}$ (b), $x=x_{\rm L}$ (c) and $x=x_{\rm R}$ (d); the laminar 
  profile $U_{\rm b}(y)\equiv y$ is indicated by a dashed-dotted line.}
\label{ob5}
\end{center}
\end{figure}

In Figure~\ref{ob5} (a) we display the profiles of $\bU_0$ and
$\bU_1$ along a streamwise line going through the center of the spot.
The dashed line corresponds to $\bU_0$ and clearly points
out the inwards character of the drift flow. In contrast, $\bU_1$
(solid line) presents a deep trough at the location of the spot. At
the spot's center where $\bU_0\simeq0$, the superposition of the
perturbation $u'=\bU_1 C y(1-y^2)$ and the base flow
$U_{\rm b}(y)\equiv y$,
shown in Figure~\ref{ob5} (b), displays the characteristic
{\it S\/} shape of the turbulent velocity profile expected for pCf.
The presence of the spot thus locally increases the wall
friction. At different positions inside the spot, where $\bU_0\ne0$
(and $\bW_0\ne0$), the full superposition $\overline U(y)=
y(1+\bU_1 C(1-y^2))+\bU_0 B (1-y^2)$ leads to asymmetric mean velocity
profiles (Fig.~\ref{ob5}(c) for point $x_{\rm L}$ and (d) for point
$x_{\rm R}$) that are reminiscent of the averaged profiles obtained by
Barkley \& Tuckerman in their simulations of the banded regime of
turbulent pCf \cite{BT07}.

\section{Generation of large scales from small scales}
\label{SIV}

The mechanism driving the quadrupolar drift flow is discussed
in terms of equations obtained by filtering from the model's
equations, as described in the Appendix. We focus on the slowly
varying quantities
$A_0=\overline{\Delta\Psi}_0$, $A_1=\overline{\Delta\Psi}_1$, and
$A_2=\overline{\Delta\Phi}_1$, driven by $B_1=-\xi\overline{\sUz\sVu}$ where
$\xi=\alpha_2(\beta+\beta'')>0$
and $B_2=\alpha_1\overline{(\sUz)^2-(\sWz)^2} +
\alpha_2\overline{(\sUu)^2-(\sWu)^2}$.
The latter quantities represent the
components of the Reynolds stress tensor \cite{Po00}
 which do not average
to zero over the 
surface of the spot ($B_1$ corresponds to the energy extracted from
the laminar flow and $B_2$ mostly to the energy contained in the
streamwise streaks).

\begin{figure}
\includegraphics[width=0.4\textwidth,clip]{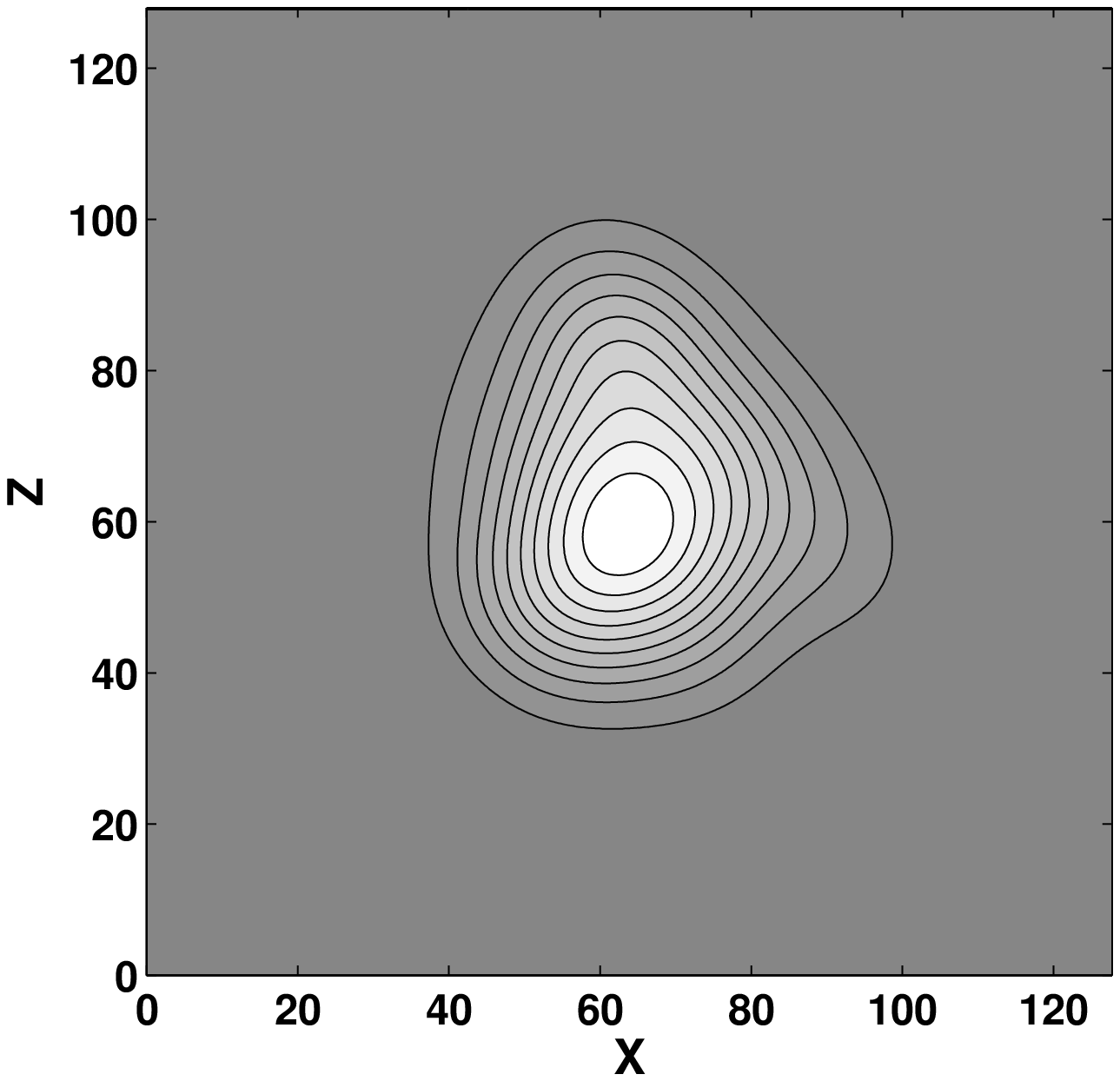}\hfill
\includegraphics[width=0.5\textwidth,clip]{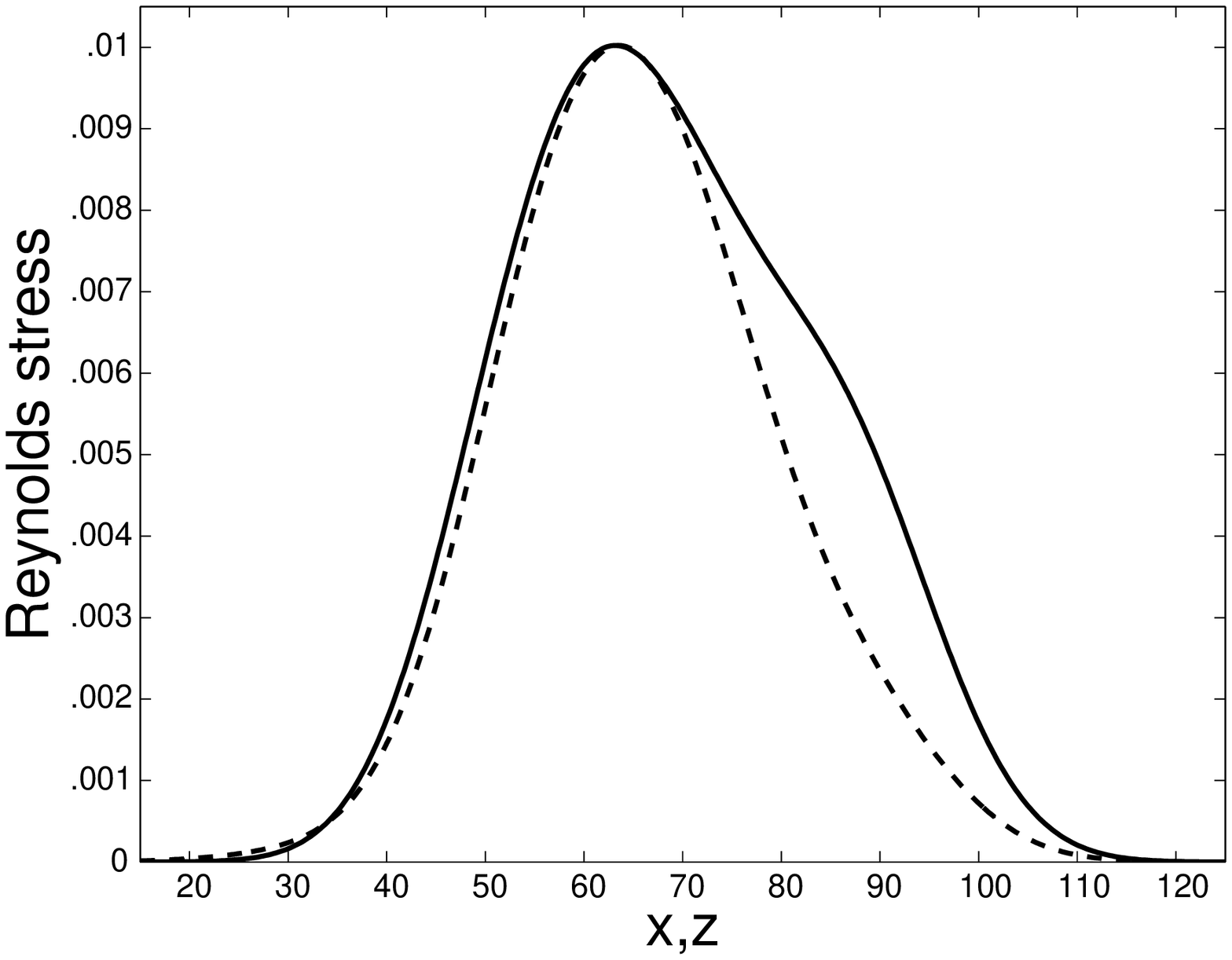}
\caption{Distribution of the averaged Reynolds stress field
$-\overline{\sUz\sVu}$ (left) and its variations along streamwise
(solid line) and spanwise (dashed line) cuts through the maximum of the
distribution taken 
as the center of the spot at $x_{\rm C}=64$, $z_{\rm C}=60$ (right).}
\label{lg03}
\end{figure}

Introducing slow variables $X$ and $Z$ whose rate
of change is inversely proportional to the width of the window that is
dragged over the data upon averaging through (\ref{E-Gauss}), one can
observe that, in the equations, the quantity $B_1$ appears with one
derivative in $X$ or $Z$ less than $B_2$, due to the fact that $B_1$
substitutes one in-plane differentiation by a cross-stream
$\mathcal O(1)$ differentiation. Further assuming that the spot is in
a quasi-steady state ($\D_t\approx0$) and that space derivatives
are negligible when compared to $\mathcal O(1)$ constants when
operating on the same quantities, at lowest significant order one can
simplify Equations (\ref{psi0pn}--\ref{phi1pn}) to read:
\BA
R^{-1}\gamma_0 A_0 &=& a_1 \left(\sfrac{3}{2}\D_Z A_2-\D_X A_1\right)\,,\label{E-A0}\\
R^{-1}\gamma_1 A_1 &=&\D_Z B_1 -a_1\D_X A_0\,,\label{E-A1}\\
R^{-1}\gamma_1 A_2 &=&-\D_X B_1\,.\label{E-A2}
\EA
The structure of this system invites one to examine the shape of the
dominant Reynolds stress contribution $B_1$ as a function of the slow
variables. Figure~\ref{lg03} displays the  averaged Reynolds stress field
associated with the small scales $-\overline{\sUz\sVu}$. As could be
anticipated the latter is positive under the spot and one can
furthermore observe its single-humped shape that, following Li \&
Widnall \cite{LW89} who developed a similar approach for spots in plane
Poiseuille flow, can be modelled as a Gaussian function of the form
$\exp[-(X^2+Z^2)/2]$. This assumption will help us to make an educated
guess about the mechanisms at work.

Considering first Equation (\ref{E-A2}), from the third equation in
(\ref{UVW1}), i.e. 
$V_1=\Delta\Phi_1/\beta$, we obtain that the contribution to $\bV_1$
generated by $B_1$ is $\sim X\exp[-(X^2+Z^2)/2]$, i.e. a pattern with
a positive hump 
for $X>0$ and a negative one for $X<0$, resembling that in
Figure~\ref{ob4} (left). This velocity component forms with $\bU_1$ a
large scale recirculation loop. As seen from the first equation in
(\ref{UVW1}), $\bU_1$ 
contains two contributions of potential and rotational origins,
respectively.
In the neighborhood of the $X$ axis, the variation of $\D_XB_1$ is
dominated by its $X$ dependence so that
$A_2=(\D_{XX}+\D_{ZZ})\overline\Phi_1\approx
\D_{XX}\overline\Phi_1=-\D_X B_1$ and, accordingly, 
$\D_X \overline \Phi_1 \sim -B_1\sim -\exp[-(X^2+Z^2)/2]$.
As to the rotational contribution $-\D_Z\overline\Psi_1$, from
(\ref{E-A1}) and forgetting the coupling with $A_0$ (which is
of higher order owing to the way it is generated from $A_1$ and
$A_2$), we have similarly 
$A_1=(\D_{XX}+\D_{ZZ})\overline\Psi_1\approx\D_{ZZ}\overline\Psi_1\sim\D_Z
B_1$, hence $-\D_Z\overline\Psi_1\sim -B_1$ so that it adds
constructively to the potential part. The resulting $\bU_1$ closes the
recirculation loop as inferred from Figure~\ref{ob4} (right).

Inserting $A_1\sim\D_Z B_1$ and 
$A_2\sim-\D_XB_1$ in (\ref{E-A0}) we obtain a right hand side in the form
$-XZ\exp[-(X^2+Z^2)/2]$ for $\Delta\overline\Psi_0$ which is
the vorticity contained in the $(U_0,W_0)$ velocity field. This field
displays four lobes with alternating signs. An approximation to the
large scale drift flow along the axes can easily be obtained.
Indeed, $\bU_0$ can be obtained from $\bU_0=-\D_Z\overline\Psi_0$ by
integrating $A_0=(\D_{XX}+\D_{ZZ})\Psi_0$ over $Z$ and neglecting
$\D_{XX}\Psi_0$ since $\Psi_0$ varies much less with $X$ than with $Z$
along the $X$ axis. We obtain
$\bU_0\sim-X\exp[-(X^2+Z^2)/2]$ which accounts for the
observed inward flow along the streamwise center-line of the spot.
The same argument can be transposed for the spanwise direction (now
$\Psi_0$ varies most rapidly in the $X$ direction, which makes
$\D_{ZZ}\Psi_0$ negligible and eases the integration over $X$), yielding
$\bW_0\sim Z\exp[-(X^2+Z^2)/2]$ which similarly accounts for
the outward flow along the spanwise center-line. Notice however that
this solution is too approximate to fulfil the continuity condition
accurately since computing $\D_X\bU_0+\D_Z\bW_0$ leaves a
residual of the form $(X^2-Z^2)\exp[-(X^2+Z^2)/2]$, though the main
contribution in $\exp[-(X^2+Z^2)/2]$ is nicely compensated
near the origin where the Gaussian is at its maximum. At any
rate the chosen shape is only a simplifying assumption.

Physically, the spot is thus characterized by a mean correction to the
base flow (represented in the model by $\bU_1<0$)
itself generated by a wall normal velocity component (here $\bV_1$)
and forming a large recirculation loop. In turn, the transport of that
mean correction (here $\bU_1 C y(1-y^2)$) by the base flow appears to
be a source term for the large scale drift flow (here $(\bU_0,\bW_0)$)
whose pattern is enslaved to its streamwise gradient, balancing
viscous forces and inertia (according to $R^{-1}\gamma_0
\bU_0+a_1\D_x\bU_1\approx0$) and expressing flow continuity
($\D_xU_0+\D_zW_0=0$).

\section{Conclusion}
\label{SV}

In this paper, we have studied the large scale structure of the flow
inside and around a turbulent spot in a transitional pCf model
focusing on the in-plane dependence of a small number of velocity
amplitudes \cite{La06}.
The approach is supported by the qualitative consistency between
previous experimental results in the transitional regime \cite{Betal98}
and our own numerical simulations of the model.

Inside the spot, we find a wide spanwise recirculation loop with
vorticity opposite in sign to that of the base flow. In particular, a patch of
streamwise correction counteracting the base flow is observed, giving
a {\it S\/} shape typical of turbulent flows to the velocity profile
inside the spot.
A reduced model (\ref{E-A1}--\ref{E-A2}) links this recirculation to
Reynolds stresses $-\overline{\sUz\sVu}$ generated by the small scale
fluctuations. 
Outside the spot, the existence of an inward-streamwise
outward-spanwise quadrupolar drift flow has been
pointed out, the origin of which is attributed to a linear coupling
with this recirculation and linked to linear momentum conservation through
(\ref{E-A0}).  By simply assuming that the region where the Reynolds
stresses contribute to the turbulent energy production (i.e.
$-\xi\overline{\sUz\sVu}>0$) is
one-humped with localized support, the main features of the large
scale flow extracted from numerical simulations by filtering are
recovered. In this approach, we only focused on the generation of large
scales by small scales but considered neither (i) the interactions
between small scales themselves nor (ii) the feedback of large scales
on small scales. Closure assumptions are clearly needed in order to
have a self-consistent theory, and especially to explain the
sustainment of turbulence within a spot, problem~(i), and its
spreading as time proceeds, problem~(ii). 

Owing to the general character of the argument leading to their
existence, one might also expect to find these large scale corrections
in and around spots developing in transitional shear flows other than
pCf for which they have already been accounted for
\cite{LJ91,SE01,Ti95}.
Evidence of their presence can indeed be obtained from Figure~12
reporting numerical work of Henningson \& Kim \cite{HK91} on plane
Poiseuille flow and from Figures~6 and~9 describing the result of
ensemble averaging of turbulent spots in boundary layer flow with
slightly adverse pressure gradient in the laboratory experiments of
Schr\"oder \& Kompenhans \cite{SK04}. Despite its limited cross-stream
resolution, our modeling of transitional plane Couette flow has thus
been shown to provide valuable explanations to previous observations,
which might call for new laboratory experiments since, besides the
theoretical challenge of understanding laminar--turbulent 
coexistence in detail, the problem of the transition to turbulence in wall
flows has a great technical importance.

\appendix
\section{Model's equations and derivation of (\ref{E-A0}--\ref{E-A2}) }

As explained in the main text, the model is obtained by projecting the
Navier--Stokes equations on the chosen basis
(\ref{pertur}--\ref{pertur3}) with velocity perturbations expanded on
the same basis. The set completing (\ref{E-cont}) and
(\ref{E-U0ns},\ref{E-NLU0}) reads: 
\begin{eqnarray*}
&&\D_t W_0 +N_{W_0}=-\D_z P_0 -a_1\D_x W_1 +R^{-1} (\Delta -\gamma_0)W_0\,,\\
&&N_{W_0}=\alpha_1 (U_0\D_x W_0+W_0 \D_z W_0)
+\alpha_2(U_1\D_xW_1+W_1\D_zW_1+\beta'V_1W_1)\,,\label{E-NW0}\\
&&\D_t U_1 +N_{U_1}=\mbox{}- \D_x P_1-a_1\D_x U_0 +R^{-1} (\Delta-\gamma_1)U_1\,,\\
&&N_{U_1}=\alpha_{2} (U_0\D_x U_1+U_1 \D_x U_0
+W_0 \D_z U_1+W_1 \D_z U_0-\beta''V_1 U_0)\,,\\
&&\D_t W_1 +N_{W_1}=\mbox{}-\D_z P_1 -a_1\D_x W_0
    + R^{-1} \left(\Delta-\gamma_1\right)W_1\,,\\
&&N_{W_1}=\alpha_{2} (U_0 \D_x W_1+U_1 \D_x W_0
+W_0\D_z W_1 +W_1 \D_z W_0-\beta''V_1 W_0)\,,\\
\label{E-V1}
&&\D_t V_1 +N_{V_1}=-\beta P_1 +R^{-1} (\Delta -\gamma_1')V_1,\\
\label{E-NV1}
&&N_{V_1}=\alpha_3(U_0\D_xV_1 +W_0\D_zV_1)\,,
\end{eqnarray*}
where $\Delta$ denotes the two-dimensional Laplacian
$\D_{xx}+\D_{zz}$. Coefficients all derive from integrals of the form 
$J_{n,m}=\int_{0}^{1} y^{n}(1-y^2)^{m} {\rm d}y
=\sum_{k=0}^{m} {k\choose m}\frac{(-1)^k}{2k+n+1}$. We have:
$a_1=1/\sqrt7$, $a_2=\sqrt{27/28}$, $\alpha_1=3\sqrt{15}/14$,
$\alpha_2=\sqrt{15}/6$, $\alpha_3=5\sqrt{15}/22$, $\gamma_0=5/2$,
$\gamma_1=21/2$,  $\gamma_1'=\beta^2$, $\beta'=\frac32\beta$, and $\beta''=\frac12\beta$.

The equations governing fields $\Psi_0$, $\Psi_1$, $\Phi_1$, from
which the velocity components derive through (\ref{UW0},\ref{UVW1}),
are obtained in the usual way by differentiating and cross-subtracting
or adding the previous equations. They read:
\BA
&&\NN(\D_t-R^{-1} (\Delta-\gamma_0)) \Delta \Psi_0\\
&&\qquad=(\D_z N_{U_0}-\D_x N_{W_0})+a_1(\sfrac{3}{2}\D_z
\Delta\Phi_1-\D_x \Delta \Psi_1)\,,\label{psi0pn}\\ 
&&\NN(\D_t-R^{-1} (\Delta-\gamma_1)) \Delta \Psi_1\\
&&\qquad=(\D_z N_{U_1}-\D_x N_{W_1})-a_1\D_x \Delta \Psi_0\,,\label{phi1pn}\\ 
&&\NN\left[\D_t(\Delta-\beta^2)-R^{-1}
(\Delta^2-2\beta^2\Delta+\gamma_1\beta^2)\right]\Delta\Phi_1\\
&&\qquad=\beta^2(\D_x N_{U_1}+\D_zN_{W_1})-\beta\Delta N_{V_1}\,.\label{psi1pn}
\EA

The introduction of averaged quantities $\tU_0$, $\tW_0$, $\tU_1$, and
$\tW_1$ in (\ref{UW0}) and 
(\ref{UVW1}) is forced by our choice of periodic boundary conditions,
otherwise the possibility of a uniform velocity correction
corresponding to linearly increasing potential/stream functions  would
be overlooked. They are governed by: 
\begin{eqnarray*}
\label{E-U0bar}
\sfrac{\rm d}{{\rm d}t}\tU_1
&=&\alpha_2(\beta+\beta''){\widetilde{U_0V_1}}
-\gamma_1R^{-1}\tU_1\,,\\
\sfrac{\rm d}{{\rm d}t}\tW_1
&=&\alpha_2(\beta+\beta''){\widetilde{W_0V_1}}
-\gamma_1R^{-1}\tW_1\,\\
\sfrac{\rm d}{{\rm d}t}\tU_0
&=&\alpha_2(\beta-\beta'){\widetilde{U_1V_1}}
- \gamma_0 R^{-1}\tU_0\,,\\
\sfrac{\rm d}{{\rm d}t}\tW_0
&=&\alpha_2(\beta-\beta'){\widetilde{W_1V_1}}
- \gamma_0 R^{-1}\tW_0\,,
\end{eqnarray*}
where the wide tildes mean averaging over the whole domain. Among this
set of equations, the first one is the most relevant since it
precisely corresponds to the expected mean flow correction. Quantity
$\alpha_2(\beta+\beta'')$ was denoted $\xi$ in the text.  

It was observed in Figure~\ref{spot_u0} that the flow within the turbulent
spot resembles developed turbulent flow, see also \cite{Hetal94,LW89}.
Accordingly,
one obtains that the only contributions to the averaged equations come
from the terms that keep a constant sign over the surface of the spot,
namely the main Reynolds stress term $-\overline{U_0V_1}$ associated
with energy extraction from the mean flow and the other terms
$\overline{U_0^2}$, $\overline{W_0^2}$, $\overline{U_1^2}$, and
$\overline{W_1^2}$. Equations (\ref{psi0pn}--\ref{phi1pn}) then reduce to:  
\BA
&&\NN\hspace*{-2em}(\D_t-R^{-1} (\Delta-\gamma_0))\Delta\overline{\Psi}_0\\
&&=\sfrac12\D_{xz}
\left(\alpha_1\overline{U_0^2-W_0^2}
       +\alpha_2\overline{U_1^2-W_1^2}\,\right)+
     a_1\left(\sfrac{3}{2}\D_z \Delta \overline{\Phi}_1 
-\D_x \Delta \overline{\Psi}_1\right),\label{psi0g}\\
&&\hspace*{-2em}(\D_t-R^{-1} (\Delta-\gamma_1)) \Delta \overline{\Psi}_1
=\D_z\left(\,\overline{-\xi U_0V_1^{\phantom{2}}}\,\right)
-a_1\D_x \Delta \overline{\Psi}_0\,,\label{psi1g}\\
&&\hspace*{-2em}\left[\D_t(\Delta-\beta^2)-R^{-1}
  (\Delta^2-2\beta^2\Delta+\gamma_1\beta^2)\right] \Delta
 \overline{\Phi}_1
=\beta^2\D_x\left(\,\overline{-\xi
    U_0V_1^{\phantom{2}}}\,\right),\label{phi1g} 
\EA
with $\xi=\alpha_2(\beta+\beta'')$. Following Li \& Widnall, we then
split the velocity components into small and large scales, i.e.
$U_0\leadsto\bU_0+U_0^{\rm s}$, etc., and only
keep the contribution to the Reynolds stresses coming from the small
scales. This leads to the same set of equations as above except that
$U_0$, $U_1$, \dots\ are replaced by their small scale parts
$U_0^{\rm s}$, $U_1^{\rm s}$, \dots.

\end{document}